\documentclass[a4paper,aps,prl,showpacs,superscriptaddress,floatfix,nofootinbib,twocolumn,bibtex]{revtex4}

\usepackage{bbold}
\usepackage{bbm}
\usepackage[pdftex]{graphicx}
\usepackage{latexsym,amsmath,verbatim}
\usepackage{color}
\usepackage{rotating}
\usepackage{multirow}
\usepackage[english]{babel}

\begin{document}

\title{A paradox in community detection
}

\author{Filippo Radicchi}
\affiliation{Center for Complex Networks and Systems Research, School of Informatics and Computing, Indiana University, Bloomington, USA}
\email{filiradi@indiana.edu.}

\date{\today}

\begin{abstract}
\noindent Recent research has shown 
that virtually all algorithms aimed at the identification
of communities in networks are affected by the same main limitation: the 
impossibility to detect communities, even when these are well-defined, if 
the average value of the difference between internal and external node degrees
does not exceed a strictly positive value, in literature known as detectability threshold. 
Here, we counterintuitively show that the value of this threshold 
is inversely proportional to the intrinsic quality of communities: 
the detection of well-defined modules is thus more difficult than the
identification of ill-defined communities.
\end{abstract}


\maketitle

\noindent Real networks are often
organized in local modules or clusters called 
communities~\cite{Girvan02, Radicchi04}.
In intuitive terms, a community
is a subgroup of nodes with a density
of internal connections larger than
the density of external links. The identification of communities 
is a crucial step for the understanding of structural
and dynamical properties of networks, and, given
the relevance of the subject, last years have witnessed
an explosion of computer algorithms 
aimed to address this challenging task~\cite{Fortunato10}.
Whereas the basic mechanisms of community detection methods can be 
diverse, several recent papers have pointed out a main limitation that affects
many identification algorithms: the existence of a so-called
detectability threshold~\cite{Rei08,Decelle10,Newman12,Radicchi13}.
To be more specific, this limitation has been
mainly investigated in the special case
of random block models composed of two subgroups
where internal and external node degrees are 
considered as independent variables extracted
from Poisson distributions with averages
equal to $\langle k_{in} \rangle$ and
$\langle k_{out} \rangle$, respectively.  
Although intuition suggests the presence of 
a well-defined community structure
for any $\Delta = \langle k_{in} \rangle - \langle k_{out} \rangle > 0$,
it has been shown that community identification algorithms are
able to detect modular structure only
when $\Delta>\Delta_c$, where the detectability 
threshold $\Delta_c = \sqrt{\langle k_{in} \rangle+ \langle k_{out} \rangle}$
is strictly larger than zero. So far, the main line of
investigation on this topic has been characterized by different
proofs of the previous result, leading therefore to the firm belief of a universal limitation
potentially affecting all community detection algorithms.
In this paper, we provide novel and contradictory
results that may cause a reconsideration of the
notion of communities in relation with the clusters
that identification algorithms effectively detect. We find
that the value of the detectability threshold increases as the ``quality''
of the communities increases. In a few words,
we counterintuitively show that the detection
of well-defined modules is more difficult than the
identification of ill-defined communities.

\

\noindent To this end, we consider a symmetric
and weighted network composed of two
subgroups of identical size $N$. 
The adjacency matrices of two
subgroups are denoted with $A$ and $B$, respectively. 
The information regarding 
external connections between nodes of the two subgroups is instead
encoded in the matrix $C$. The adjacency matrix $G$
of the entire network can be thus written in the block form
\begin{equation}
G = \left(
\begin{array}{cc}
A & C
\\
C^T & B
\end{array}
\right) \;.
\label{eq:adj}
\end{equation}
$A$, $B$ and $C$ are square
matrices of dimensions $N \times N$, and
$G$ is a symmetric matrix of dimensions $2N \times 2N$.
For simplicity, we concentrate our attention on the 
ensemble of graphs generated according to
the following procedure. Both
subgroups have the same internal degree sequence 
$k_{in} = \{k_1^{in}, \ldots, k_N^{in}\}$, and the 
same external degree 
sequence $k_{out} = \{k_1^{out}, \ldots, k_N^{out}\}$. 
Apart for such constraints, the graph is completely random in the sense 
that connections among nodes are randomly drawn
with the only prescription of preserving the {\it a priori} given 
internal and external degree sequences~\cite{Molloy95}.
In the construction of our ensemble, we limit to
the case in which the subgroups are composed of
a single connected component, but 
we allow for the eventual presence of self-loops and
multiple connections among the same pairs of nodes.
Finally, in order to provide results
easily comparable with those obtained in previous
works, we focus on the case in which the 
entries of the degree sequences $k_{in}$
and $k_{out}$ are random variates
extracted from Poisson distributions with 
averages $\langle k_{in} \rangle$ and $\langle k_{out} \rangle$, respectively,
and the sum of these average values is subjected to the constraint 
$\Sigma = \langle k_{in} \rangle +\langle k_{out} \rangle$.

\

\noindent From now on, we analyze the ability to reveal 
the presence of the two subgroups
of the second smallest eigenvector of the normalized 
laplacian $\mathcal{L}$ associated to the adjacency matrix $G$. 
This method is very popular in graph clustering,
and represents a way to find the bipartition
corresponding to the minimum of the so-called
normalized cut of the graph~\cite{Shi97}.
More importantly, this approach is essentially equivalent to other 
spectral clustering methodologies
(modularity maximization and statistical inference)
as shown in a recent work by Newman~\cite{Newman13}, so that
the following results are potentially valid for a more general
class of community identification algorithms.
The normalized laplacian $\mathcal{L}$ associated
to the adjacency matrix $G$ is defined as
\begin{equation}
\mathcal{L} = \mathbbm{1} - D^{-1/2} G D^{-1/2} \;,
\label{eq:lap}
\end{equation}
where $\mathbbm{1}$ is the identity matrix, and
$D$ is a diagonal matrix whose
diagonal elements are equal to the node degrees~\cite{Chung_book}.
Let us denote with $(\nu_2, \mathbf{v}_2)$ the second smallest
eigenpair of $\mathcal{L}$. 
The bipartition corresponding to minimum normalized cut 
is determined on the basis of the signs
of the components of $\mathbf{v}_2$ by placing
each node in one of two modules if
its corresponding component in $\mathbf{v}_2$ 
is negative or positive~\cite{Shi97}. 
For our purposes, it is thus important to
determine whether these modules correspond to the two
pre-imposed subgroups of our network block model or not.

\begin{figure}
\begin{center}
\includegraphics[width=0.45\textwidth]{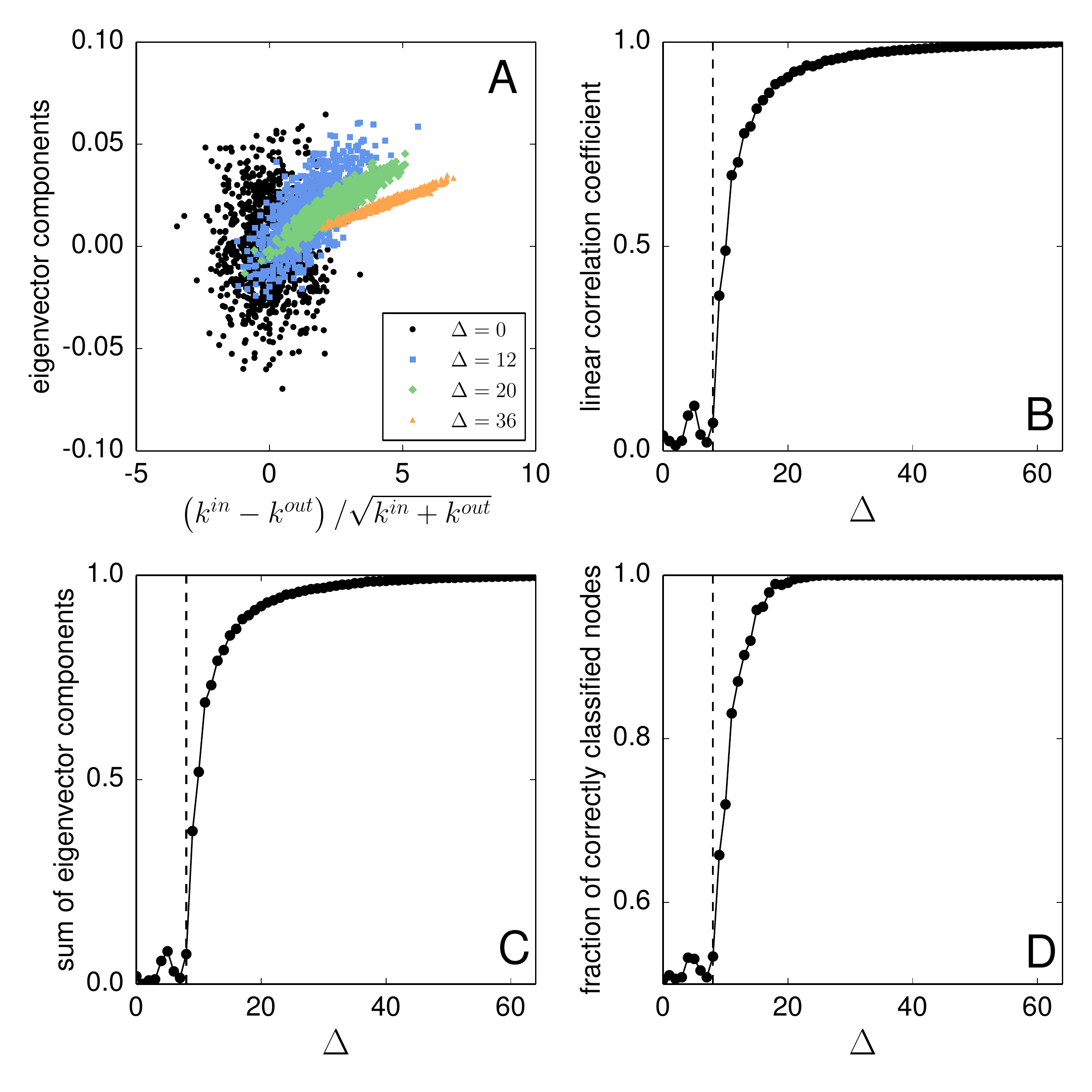}
\end{center}
\caption{Analysis of networks composed of two subgroups of
size $N=1024$. Average internal and external degrees
are chosen such that their sum is $\Sigma = 
\langle k_{in} \rangle + \langle k_{out} \rangle =64$.
In- and out-degree sequences are neutrally correlated
in the sense that the correlation term of Eq.~(\ref{eq:corr})
reads $\xi = \xi_{rand} = \langle k_{in} \rangle \langle k_{out} \rangle$. 
{\bf A} Numerical validation of the ansatz of Eq.~(\ref{eq:eivec}).
Please note that here we are showing only the components
of the eigenvector $\mathbf{v}_2$ that
correspond to nodes of one of the two subgroups.
Different colors and symbols correspond
to different choices of the difference
$\Delta = \langle k_{in} \rangle - \langle k_{out} \rangle$.
{\bf B} Linear correlation coefficient
between the r.h.s. of Eq.~(\ref{eq:eivec})
and the numerical estimation of the
components of $\mathbf{v}_2$ as a function of $\Delta$.
{\bf C} Sum of the components of $\mathbf{v}_2$ as a function of $\Delta$.
Only the components corresponding to nodes of a single group are considered.
This sum is further divided by $\sqrt{N/2}$ in order
to obtain a number between zero and one.
{\bf D} Fraction of correctly classified nodes as a function of $\Delta$.
In panels B, C and D, the dashed lines stand for the prediction
of the detectability threshold 
$\Delta_c = \sqrt{\Sigma}$ [Eq.~(\ref{eq:thresh})].
}
\end{figure}

\

\noindent Following the same lines of the approach 
described in~\cite{Radicchi13, Radicchi_new}, 
it is possible to show that the presence of modular structure is 
revealed by an eigenvector
whose $i$-th component behaves, on average, as
\begin{equation}
v^*_i = q_i \, \frac{k_i^{in}-k_i^{out}}{\sqrt{k_i^{in}+k_i^{out}}} \; ,  
\label{eq:eivec}
\end{equation}
with $q_i = \pm q$ if node $i$ belongs to the first subgroup, 
$q_i = \mp q$ otherwise, 
and $q$ proper normalization constant. 
The validity of the previous approximation 
can be determined by a direct comparison between the components of the
second smallest eigenvector $\mathbf{v}_2$ 
obtained in numerical experiments
and the r.h.s. of Eq.~(\ref{eq:eivec}), as shown for example
in Fig.~1A. Using  $\mathbf{v}^*$ as ansatz in the 
eigenvalue problem $\mathcal{L} \mathbf{v}^* = \nu^* \mathbf{v}^*$, 
it is possible to further determine the associated eigenvalue as
\begin{equation}
\nu^* = 1 - \frac{\langle k_{in}^2 \rangle + \langle k_{out}^2 \rangle - 2 \xi}{\langle k_{in}^2 \rangle - \langle k_{out}^2 \rangle} \; .
\label{eq:eigen}
\end{equation}
In Eq.~(\ref{eq:eigen}), $\langle k_{in}^2 \rangle$ and $\langle k_{out}^2 \rangle$
are the second moments of the in- and out-degree distributions of 
the two groups (for
Poisson distributions, we have
$\langle k_{in}^2 \rangle = \langle k_{in} \rangle^2 + \langle k_{in} \rangle$
and $\langle k_{out}^2 \rangle = \langle k_{out} \rangle^2 + \langle k_{out} \rangle$), while $\xi$, defined as
\begin{equation}
\xi =  \sum_{k^{in}, k^{out}} k^{in} k^{out}\, P(k^{in}, k^{out}) \;,
\label{eq:corr}
\end{equation}
with $P(k^{in}, k^{out})$ probability to find a node
with internal and external degrees
respectively equal to $k^{in}$ and $k^{out}$,
represents the correlation between the number of internal and external
connections of the nodes in the network.
Please note that the correlation term $\xi$ can be regulated 
by simply permuting the entries of the in- and out-degree sequences, 
and thus without changing the correspondent distributions.

\

\noindent In order to determine
the detectability threshold, we have to understand
whether $\nu^*$ effectively represents
the second smallest eigenvalue $\nu_2$ of $\mathcal{L}$.
If this is true then
communities are detectable in terms of the components
of the eigenvector $\mathbf{v}^*=\mathbf{v}_2$ defined
in Eq.~(\ref{eq:eivec}), otherwise they are not.
The term of comparison of $\nu^*$ is given
by the expected value $\nu_{rand}$ of the
second smallest eigenvalue of the normalized laplacian
of a random network
with the same average degree 
$\Sigma = \langle k_{in} \rangle + \langle k_{out} \rangle$, whose
value can be estimated, thanks to the
predictions by Chung and collaborators~\cite{Chung03, Chung03a}, as
\begin{equation}
\nu_{rand} \simeq 1 - \frac{2}{\sqrt{\langle k_{in} \rangle + \langle k_{out} \rangle}} \; .
\label{eq:chung}
\end{equation}
If $\nu^* < \nu_{rand}$, this means that
$\nu_2 = \nu^*$ and the conditions
to reveal the presence of two subgroups are satisfied. If instead
$\nu^* > \nu_{rand}$, then $\nu_2=\nu_{rand}$, and the two subgroups
are not detectable by means of the components of the second
smallest eigenvector of the normalized laplacian.
The condition $\nu^* = \nu_{rand}$
determines the detectability threshold $\Delta_c$, i.e., the minimal value
of the difference $\Delta = \langle k_{in} \rangle - \langle k_{out} \rangle$ for which
$\nu^*$ equals the typical second smallest eigenvalue 
$\nu_{rand}$ of a random graph
with identical average degree $\Sigma$. 
As a direct
comparison between Eqs.~(\ref{eq:eigen}) and~(\ref{eq:chung}) reveals, 
the value of $\Delta_c$ does not depend only on the average values of internal and external
degrees, but also on the correlation
term $\xi$ defined in Eq.~(\ref{eq:corr}). 
In the following, we consider  
three special cases.

\begin{figure}
\begin{center}
\includegraphics[width=0.45\textwidth]{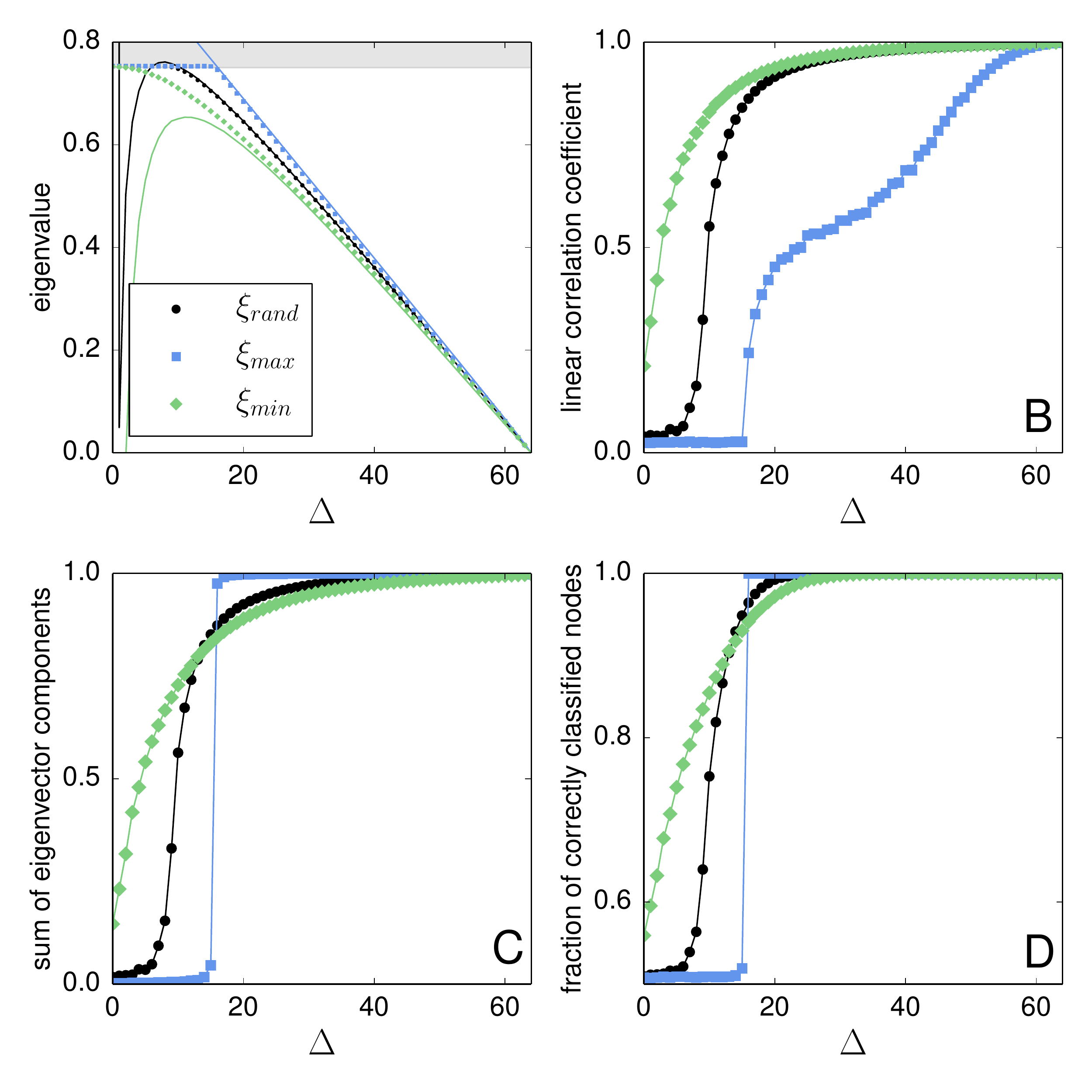}
\end{center}
\caption{Analysis of networks composed of two subgroups of
size $N=1024$. Average internal and external degrees
are chosen such that their sum is $\Sigma =
 \langle k_{in} \rangle + \langle k_{out} \rangle =64$.
Different colors and symbols correspond to
the three different values of the correlation
term $\xi$ described in the text: 
$\xi_{min}$ (green diamonds), 
$\xi_{rand}$ (black circles), $\xi_{max}$ (blue squares).
All numerical results shown here correspond
to average values obtained in 100 realizations of the
network models.
{\bf A} Numerical estimation of the second smallest
eigenvalue $\nu_2$ (symbols). Full lines are obtained with
Eq.~(\ref{eq:eigen}). The gray area delimits the expected 
spectral radius of the normalized
laplacian of a random graph with identical average degree
[i.e., Eq.~(\ref{eq:chung})].
{\bf B} Linear correlation coefficient between
the components of $\mathbf{v}_2$ and the r.h.s. of
Eq.~(\ref{eq:eivec}) as a function 
of $\Delta = \langle k_{in} \rangle - \langle k_{out} \rangle$.
Only the components of $\mathbf{v}_2$ corresponding to
nodes of one of the subgroups are considered.
{\bf C} Sum of the components of $\mathbf{v}_2$ 
corresponding to nodes of one of the two subgroups as a function
of $\Delta$. To generate numbers between zero and one, this
sum is divided by the constant factor $\sqrt{N/2}$.
{\bf D} Fraction of nodes correctly classified as a function of $\Delta$.
}
\end{figure}

\

\noindent {\it Neutrally correlated degree sequences}. This is
the case usually considered in literature for the determination
of the detectability threshold. In- and out-degrees
are considered as independent variables, so that the correlation
term reads $\xi = \xi_{rand} = \langle k_{in} \rangle \langle k_{out} \rangle$.
By equating the r.h.s. of Eqs.~(\ref{eq:eigen}) and~(\ref{eq:chung}), 
using the definition of the second moments of
Poisson distributions, and
assuming $\Sigma \gg 1$, we recover the well known result
\begin{equation}
\Delta_c = \sqrt{\Sigma} \; .
\label{eq:thresh}
\end{equation}
Such prediction is in perfect agreement with
the results of numerical experiments as reported in Figs.~1
and~2. To monitor the transition between the 
undetectable and detectable regimes, we use 
three different order parameters: (i) The 
absolute value of the linear correlation
coefficient between the components of the eigenvector
$\mathbf{v}_2$ (numerically estimated) and the
r.h.s. of Eq.~(\ref{eq:eivec}). Please note that in the
evaluation of the linear correlation coefficient we consider
only the first $N$ components of $\mathbf{v}_2$, i.e.,
the components corresponding to nodes inside the first
subgroup. (ii) The absolute value of the sum of the components
of the eigenvector $\mathbf{v}_2$ corresponding to nodes
within the first subgroup. For convenience, this sum is further 
divided by the factor $\sqrt{N/2}$ to obtain numbers
between zero and one. (iii) Fraction of nodes correctly classified.
We consider the classification provided by the signs
of the components of the eigenvector $\mathbf{v}_2$
and compare it with the pre-imposed division in two subgroups.
All the three order parameters clearly show
the presence of a transition, as a function of $\Delta$,
between a regime in which the modules
identified by the spectral algorithm do not correspond to the 
pre-imposed subgroups, and a regime
in which the detected communities coincide instead with them.
More importantly, the value of $\Delta$ for which such
transition occurs is well approximated by Eq.~(\ref{eq:thresh}).
In intuitive terms, one could try to justify the 
detectability threshold of Eq.~(\ref{eq:thresh})
with the following argument. The fact
that internal and external degrees are independent variables
allows for the presence of nodes whose
internal degree is lower than their external number
connections even for 
$\Delta 
> 0$.
The assignment to a specific group of nodes with more external than internal
connections is technically incorrect, and
the second smallest eigenvector of $\mathcal{L}$ provides indeed the
right answer by ``misplacing'' these nodes.
Actually, the difference between internal and external node degrees 
is a random variate that obeys the so-called Skellam
distribution with mean equal to $\Delta$ and standard deviation
equal to $\sqrt{\Sigma}$~\cite{Skellam46}. Thus, the second smallest
eigenvector of $\mathcal{L}$ starts to classify nodes
in the pre-imposed subgroups only when the average value of 
the difference between internal and external degrees
is larger than the typical variability of the same variable.
This straightforward interpretation is, however, contradicted
by the following cases.

\

\noindent {\it Positively correlated degree sequences}.
If we sort the entries of both internal and external degree
sequences in ascending (descending) order, then nodes with
high internal degree have high external degree, and {\it vice versa}.
The re-arrangement inequality 
tells us that the correlation term reads $\xi = \xi_{max}$,
being 
$\xi_{max}$ the maximum
value of $\xi$ that can be reached for fixed entries in the
degree sequences~\cite{Dar73}. Please note that in this
case the subgroups in our graph are ``well-defined''
communities. Apart for extreme cases,
we expect in fact that each node in the network have more internal
than external connections for every choice of
$\Delta
> 0$.
According to the intuitive argument used to justify Eq.~(\ref{eq:thresh}),
we expect to see a detectability threshold not only smaller
than the one given in Eq.~(\ref{eq:thresh}), but also 
very close to zero. Contrary to intuition, however, 
the detectability threshold becomes larger (see Fig.~2). 
Although we cannot provide an exact estimation of $\Delta_c$
because we are not able to analytically determine 
the value of $\xi_{max}$, we can anyway see from our Eq.~(\ref{eq:eigen}) 
why this counterintuitive behavior is indeed expected.
As $\xi$ increases in fact, the r.h.s. of Eq.~(\ref{eq:eigen})
gets closer to the
typical second smallest eigenvalue of a random graph
with average degree $\Sigma$ [i.e., Eq.~(\ref{eq:chung})].
We thus require large values of $\Delta$ in order
to make $\nu^*$ sufficiently small.

\

\noindent {\it Negatively correlated degree sequences}.
If we sort the entries of the internal degree
sequence in ascending (descending) order, 
and those of the external degree sequence in
descending (ascending) order, then nodes with
high internal degree have low external degree, and {\it vice versa}.
In this case, the re-arrangement inequality 
states that the correlation term reads $\xi = \xi_{min}$,
being 
$\xi_{min}$
the minimal value of $\xi$ that can be reached for fixed entries in the
degree sequences~\cite{Dar73}.
This is the antithetic case of the one just presented for
positively correlated degree sequences: even for
large values of $\Delta$, the pre-imposed subgroups
are ``ill-defined'' communities, in the sense that it is very likely
to find nodes with external degree
larger their internal degree. We should thus expect a very large
detectability threshold, but again this is
expectation is violated. 
The r.h.s. of Eq.~(\ref{eq:eigen}) decreases as $\xi$ decreases,
so that we require smaller values of $\Delta$ to make
$\nu^*$ smaller than $\nu_{rand}$.
Although in this case, our prediction of Eq.~(\ref{eq:eigen})
fails to correctly describe the behavior of the second smallest
eigenvalue of the normalized laplacian (see Fig.~2),
our numerical computations show that $\Delta_c = 0$, and thus
communities are always detectable if
$\langle k_{in} \rangle > \langle k_{out} \rangle$.

\begin{figure}
\begin{center}
\includegraphics[width=0.45\textwidth]{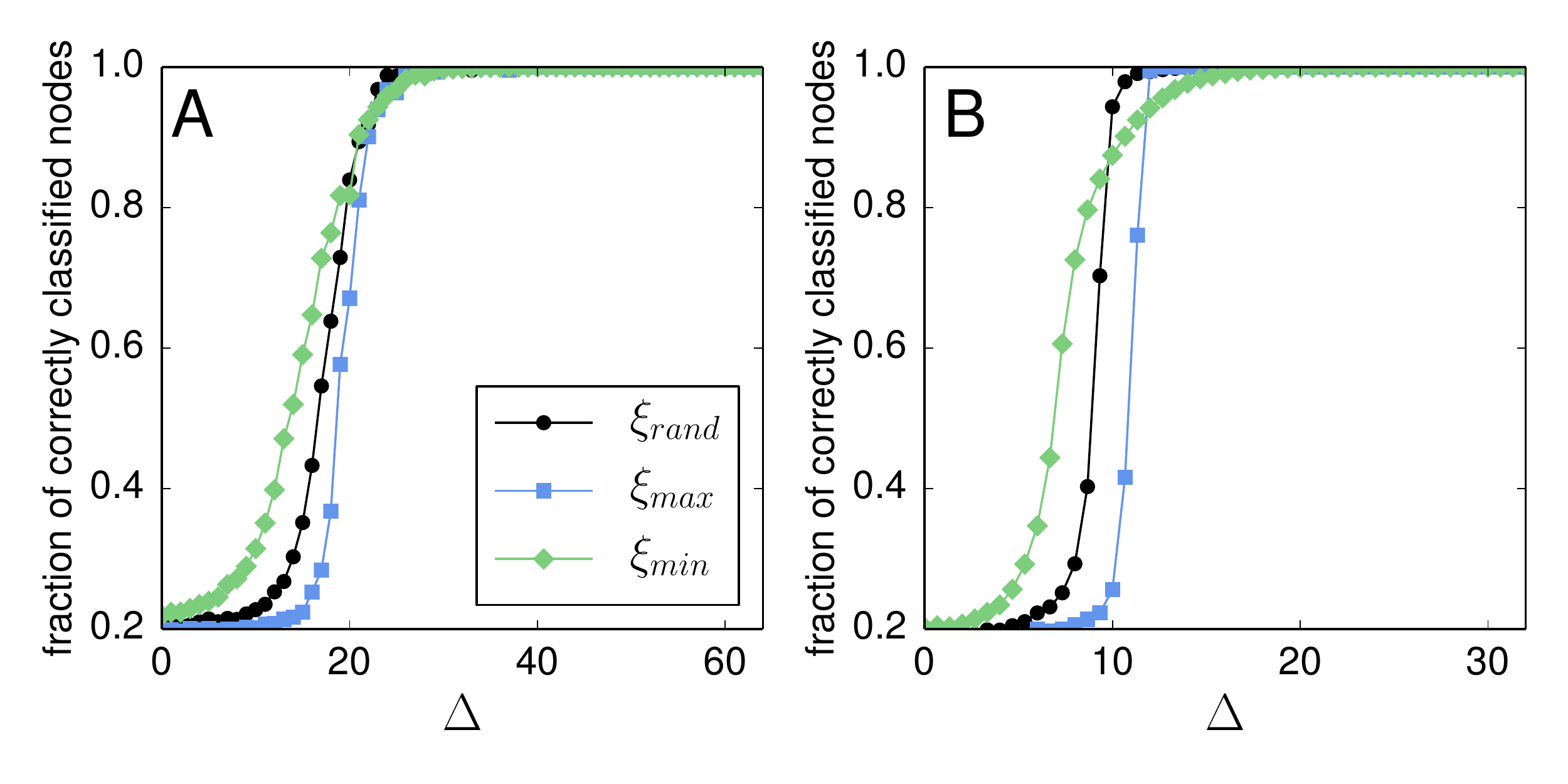}
\end{center}
\caption{
Application of the algorithm by Blondel {\it et al.}
to the detection of communities in our network block 
model~\cite{Blondel08}. We plot the fraction of nodes
that the algorithm correctly classifies as a function
of $\Delta = \langle k^{in} \rangle - \langle k^{out} \rangle$. 
Different colors and symbols correspond
to different choices of $\xi$. Each point 
represents the average performance
of the algorithm in
$100$ realizations of the model.
{\bf A} Block model composed of two subgroups of
size $N=1024$ and average degree 
$\Sigma =  \langle k^{in} \rangle + \langle k^{out} \rangle = 64$.
{\bf B} Block model composed of four subgroups of
size $N=512$ and average degree 
$\Sigma = \langle k^{in} \rangle + 3 \langle k^{out} \rangle = 32$.
}
\end{figure}

\

\noindent To summarize,
the ability of algorithms to identify communities in networks 
is dependent on their intrinsic quality.
We use the word ``quality'' in relation with the intuitive notion 
of communities, and thus
on the comparison between the number of internal and external connections
at the node level. A ``well-defined'' 
community is a subgroup 
in which each node has more connections with
other nodes inside the community than with vertices
outside of it, whereas an ``ill-defined'' community
is a subgroup in which many nodes have more
connections outside than inside the community.
Intuitively, we should expect that good communities are
less challenging to be identified than bad communities, 
but our results show the exact contrary. 
This situation is paradoxical
in the sense that community detection algorithms are 
developed with the aim
of detecting particular kind of substructures,
but, in reality, they are better suited
to detect groups of nodes that do not
fully respect the intuitive definition of communities.
One could argue that our results are not general enough 
because they are based
on the performance of a
special type of algorithm that is ``forced'' to classify
nodes in two modules, and the random block model on which
this algorithm is applied is also composed of two subgroups.
These, however, do not seem
serious limitations. In Fig.~3, we report the
results obtained with the popular community detection
algorithm by Blondel {\it et al.}~\cite{Blondel08}. 
It is important to stress that this algorithm does not use 
any previous knowledge regarding the number of subgroups pre-imposed
in our block model. Its performances
are in line with what shown so far for spectral
algorithms: the detection of well-defined communities is more
challenging than the identification of ill-defined
communities. This is not only valid for the case of 
two subgroups (Fig.~3A), but applies also to higher
numbers of subgroups (Fig.~3B).
Although this additional analysis gives a
more general character to our results, we do not exclude the
presence of other factors that can have an influence
on the behavior of the detectability threshold, as for example
heterogeneous community sizes and/or node degrees. 
Accounting for other relevant effects surely represents 
the next step towards a deeper understanding of this 
appealing and important problem.
Also, in the search of possible resolutions
of the paradoxical situation illustrated in this paper, our results indicates a 
promising direction to explore: the numerator
$\langle (k_{in} - k_{out})^2 \rangle = 
\langle k_{in}^2 \rangle + \langle k_{out}^2 \rangle - 2 \xi$ 
of the fraction appearing in
Eq.~(\ref{eq:eigen}) suggests that
the second moment of the distribution of the variables $k_{in}-k_{out}$ 
play a role as important as its first moment
in the determination of the
detectability threshold. We believe that this information can be used
to redesign the mechanisms at the basis of 
community detection algorithms.

\

\noindent
As a final remark, we would like to stress 
that the relevance of our results is not only theoretical, 
but also practical. Consider for example the case of artificial graphs
used to test the performance of community detection
algorithms~\cite{danon2005comparing, lancichinetti2009community}.
These are typically constructed on the
basis of random block models 
very similar to the one considered in this paper. However, depending
on the particular recipe used to construct them, the correlation 
between internal and external degrees at the node level 
can be very different. As paradigmatic examples,
consider the two benchmark graphs that are widely adopted
for testing the performance of community detection algorithms:
the Girvan-Newman (GN) and the  
Lancichinetti-Fortunato-Radicchi (LFR) 
benchmark graphs~\cite{Girvan02, Lancichinetti08}.
The LFR model is generally considered the 
extension of the GN model to heterogeneous node degrees
and community sizes, but this is
not totally correct. In the GN model, internal and external degrees
are neutrally correlated; on contrary in the LFR graphs, 
internal and external degrees are positively correlated.
Comparisons in the performance of community
identification methods 
across these two benchmark models are thus
not possible, and this is a fundamental
consideration to take into account when choosing the best
algorithm to detect {\it a priori} unknown clusters in real networks.

\begin{acknowledgments}
\noindent The author thanks A. Lancichinetti for
comments on the manuscript.
\end{acknowledgments}

\bibliography{paradox}

\end{document}